\documentclass[aps,prl,floatfix,twocolumn,superscriptaddress]{revtex4-1}
\pdfoutput=1
\usepackage{amsmath}
\usepackage{amssymb}
\usepackage{amstext}
\usepackage{amsopn}
\usepackage{amsfonts}
\usepackage{amsxtra}
\usepackage{graphicx}
\usepackage{float}
\usepackage{bm}
\usepackage{pifont}
\usepackage{hyperref}
\begin{document}

\title{Optical conductivity of nodal metals}
\author{C. C. Homes}
\email{homes@bnl.gov}
\affiliation{Condensed Matter Physics and Materials Science Department,
  Brookhaven National Laboratory, Upton, New York 11973, USA}%
\author{J. J. Tu}
\email{jtu@sci.ccny.cuny.edu}
\author{J. Li}
\affiliation{Department of Physics, The City College of New York, New York, New York 10031, USA}
\author{G. D. Gu}
\affiliation{Condensed Matter Physics and Materials Science Department,
  Brookhaven National Laboratory, Upton, New York 11973, USA}%
\author{A. Akrap}
\affiliation{\'{E}cole de Physique, Universit\'{e} de Gen\`{e}ve, CH-1211 Gen\`{e}ve 4, Switzerland}
\date{\today}

\begin{abstract}
Fermi liquid theory is remarkably successful in describing the transport and
optical properties of metals; at frequencies higher than the scattering rate,
the optical conductivity adopts the well-known power law behavior
$\sigma_1(\omega) \propto \omega^{-2}$.
We have observed an unusual non-Fermi liquid response
$\sigma_1(\omega) \propto \omega^{-1\pm 0.2}$ in the ground states of several cuprate
and iron-based materials which undergo electronic or magnetic phase transitions
resulting in dramatically reduced or nodal Fermi surfaces.  The identification of an
inverse (or fractional)
power-law behavior in the residual optical conductivity now permits the removal of this
contribution, revealing the direct transitions across the gap and allowing the
nature of the electron-boson coupling to be probed.
The non-Fermi liquid behavior in these systems may be the result of a common
Fermi surface topology of Dirac cone-like features in the electronic dispersion.
\end{abstract}

\maketitle
%

%
%
%
\noindent In a Fermi liquid, the complex conductivity $\tilde\sigma = \sigma_1 +i\sigma_2$
can be expressed through the generalized Drude model,
$\tilde\sigma(\omega) = \left( \omega_p^2/60 \right) / \left\{ 1/\tau(\omega) -
i\omega \left[1+\lambda(\omega)\right] \right\}$
%
%
(in units of $\Omega^{-1}$cm$^{-1}$), where $\omega_{p}^2 = 4\pi ne^2/m_b$, $1/\tau(\omega)$
and $1+\lambda(\omega) = m^\ast(\omega)/m_b$ are the plasma frequency, frequency-dependent
scattering rate and mass enhancement, respectively, where $n$ is a carrier concentration and $m_b$
is the band mass.  At low-temperature the scattering rate will vary quadratically
with frequency and temperature, $1/\tau(\omega,T) = 1/\tau_0 + A \left[ (\hbar\omega)^2 +
(2\pi k_{\rm B}T)^2 \right]$, where $A$ is a constant that varies with the
material \cite{gurzhi65,nagel12}.  In the frequency domain, for $\omega\tau \ll 1$ the
conductivity varies slowly, but for $\omega\tau \gg 1$ the conductivity
adopts a power-law behavior, $\sigma_1 \propto \omega^{-2}$; however, deviations from
this behavior may be observed in strongly-correlated electronic
systems \cite{vdmarel99,dodge00,orenstein00,dordevic06,berthod13}.
%

%
%
\section*{Results}
The temperature dependence of the optical conductivity of optimally-doped
Bi$_2$Sr$_2$CaCu$_2$O$_{8+\delta}$, one of the most thoroughly studied cuprate
high-temperature superconductors \cite{basov05}, is shown versus wave number
(photon energy) in a log-log plot in Fig.~1a for light polarized along
the crystallographic {\em a} axis \cite{tu02}.  Just above $T_c$ it may be
argued that the optical properties are consistent with those of a Fermi liquid
(see Supplementary Information and Fig.~S1 online for a discussion
of different models for the optical conductivity and the frequency-dependent
scattering rate); this statement is in keeping with the proposed phase diagram
for the high-temperature superconductors \cite{broun08}.
Below $T_c$ there is a rapid reduction of the low-frequency conductivity or
spectral weight, which is defined as the area under the conductivity curve; this
`missing spectral weight' is the optical signature for the formation of a
superconducting condensate \cite{basov05}.  However, even down to the lowest
measured temperature there is still a significant amount of low-frequency residual
conductivity \cite{buckley01}. This is because, unlike a conventional {\em s}-wave
superconductor in which the entire Fermi surface is completely gapped below $T_c$,
the cuprate materials have a momentum-dependent {\em d}-wave gap that contains
nodes \cite{harlingen95,damascelli03}, $\Delta(\mathbf{k}) = \Delta_0 \left[ \cos(k_x a)
- \cos(k_y a) \right]$, where $\Delta_0$ is the gap maximum.
%
%
%
\begin{figure*}
\centerline{\includegraphics[angle=270,width=6.5in]{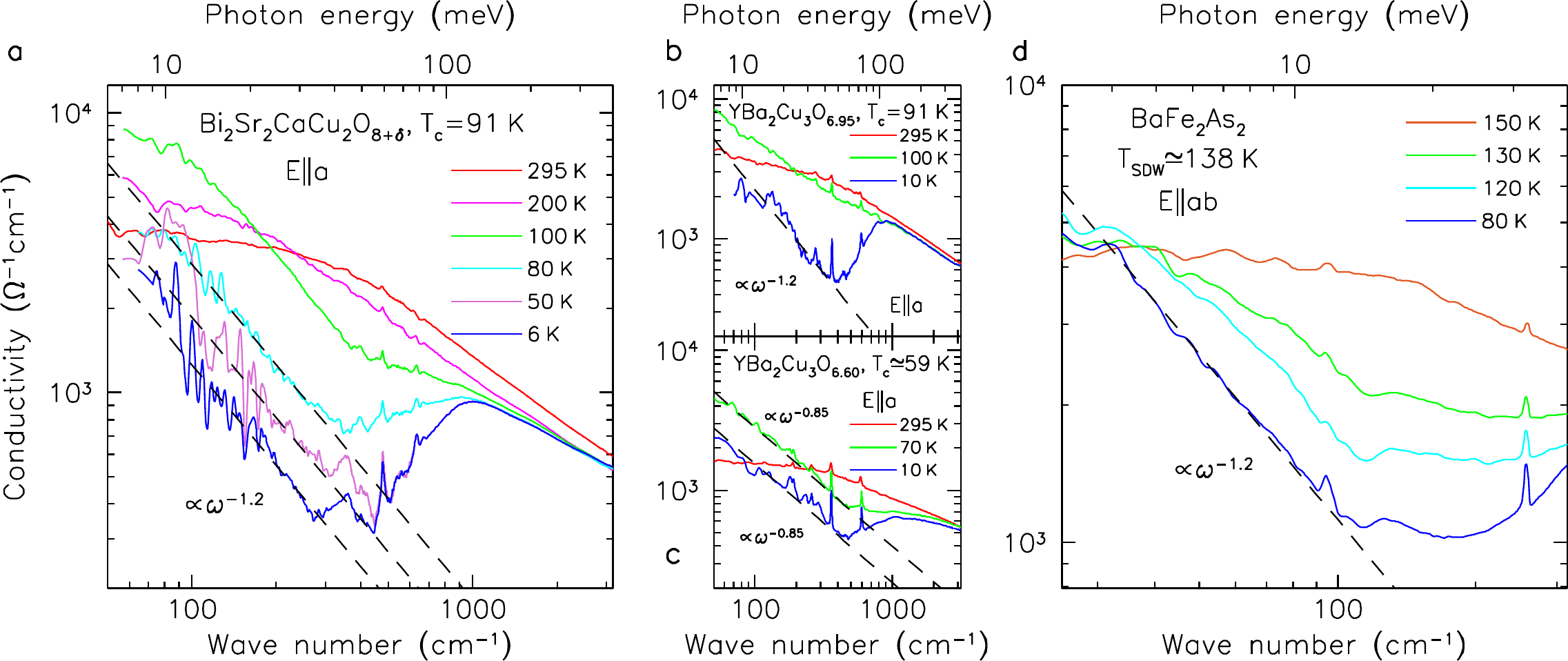}}
\vspace*{0.4cm}
\begin{trivlist}
\item
\boldmath
\noindent{\bf\textsf{Figure 1} $|$ The optical conductivity of some quantum materials.}
\unboldmath
{\bf a}, The temperature dependence of the optical conductivity versus wave number
(photon energy) for optimally-doped Bi$_2$Sr$_2$CaCu$_2$O$_{8+\delta}$
($T_c = 91$~K) for light polarized along the crystallographic {\em a} axis.
At low frequency just above $T_c$ the material may be cautiously described as a
Fermi liquid.  For all the temperatures measured below $T_c$ the residual
conductivity from the unpaired quasiparticles follows the same non-Fermi liquid
fractional power law, $\sigma_1(\omega) \propto \omega^{-1.2}$.
{\bf b}, The plot for optimally-doped YBa$_2$Cu$_3$O$_{6.95}$ ($T_c = 91$~K),
for light polarized along the {\em a} axis, illustrating the fractional power
law below $T_c$.
{\bf c}, The plot for underdoped YBa$_2$Cu$_3$O$_{6.60}$ ($T_c \simeq 59$~K),
for light polarized along the {\em a} axis, illustrating an identical (non-Fermi
liquid) fractional power-law behavior in the normal (pseudogap) and superconducting
states.
{\bf d}, The plot for the BaFe$_2$As$_2$ ($T_{\rm SDW} = 138$~K), for light polarized
in the {\em a-b} planes.  Below $T_{SDW}$ the fractional power law
$\sigma_1 \propto \omega^{-1.2}$ is again observed.
\end{trivlist}
\end{figure*}

The presence of nodes allows pair-breaking out of the superconducting state
resulting in unpaired nodal quasiparticles \cite{lee93}.  For low photon energies
($\hbar\omega \ll 2\Delta_0$) only the nodal structure of the {\em d}-wave gap is
probed and the Fermi surface topology is similar to that of the Dirac cone observed
in graphene and other quantum materials \cite{geim07,moore10}.  The rapid collapse of
the quasiparticle scattering rate \cite{shibauchi96} below $T_c$ indicates that in the
far-infrared region $\omega\tau \gg 1$, so $\sigma_1 \propto \omega^{-2}$ should be
clearly revealed.  Surprisingly, what is observed instead is that below $T_c$
low-frequency residual optical conductivity forms a family of lines with the
same non-Fermi liquid fractional power law behavior $\sigma_1 \propto \omega^{-1.2}$;
in metallic systems at low-frequency where $\sigma_1 \gg \sigma_2$, this is
approximately equivalent to the scattering rate having a fractional power law
behavior $1/\tau \propto \omega^{1.2}$.
%
%
Another family of cuprates that has been extensively investigated are the
YBa$_2$Cu$_3$O$_{6+y}$ materials.  The optical conductivity of optimally-doped
YBa$_2$Cu$_3$O$_{6.95}$ is shown in Fig.~1b for light polarized along the {\em a}
axis; this crystallographic axis is transverse to the copper-oxygen chains and
should therefore probe the dynamics of only the copper-oxygen planes \cite{homes99}.
Well below $T_c$, the observed power law for the residual conductivity
$\sigma_1 \propto \omega^{-1.2}$ is identical to the response observed in
optimally-doped Bi$_2$Sr$_2$CaCu$_2$O$_{8+\delta}$.
%
%
The underdoped YBa$_2$Cu$_3$O$_{6.60}$ sample is of particular interest due to the
formation of a pseudogap in the normal state \cite{timusk99} and the commensurate
reduction of the Fermi surface around the nodal regions, a condition that has been
referred to as a `nodal metal' \cite{ando01,lee05}.  The optical conductivity
for this material is shown in Fig.~1c for light polarized along the {\em a} axis.
Just above $T_c$ in the normal state the low-frequency optical conductivity may be
described using a non-Fermi liquid power-law, $\sigma_1 \propto \omega^{-0.85}$;
however, what is fascinating is that well below $T_c$ the response of the unpaired
quasiparticles displays the identical fractional power law.
%
%
This indicates the (unpaired) quasiparticles appear to behave the same way regardless
of whether it is the pseudogap that results in the reduction of a large Fermi surface
to a small arc or pocket \cite{kanigel06}, or the formation of a {\em d}-wave
superconducting energy gap resulting in nodes.
This non-Fermi liquid power-law behavior in the underdoped material has been previously
observed in the microwave region \cite{turner03}; however, in that work the exponent is
considerably larger, $\sigma_1 \propto \omega^{-1.45}$.  The most likely source for this
disagreement is the fact that the microwave experiments are done in the $\omega\tau
\sim 1$ region, while the optical work was performed in the $\omega\tau \gg 1$
limit, suggesting that the relaxation processes in these two regimes may be different.
Surprisingly, recent results on the single-layer, underdoped cuprate HgBa$_2$CuO$_{4+\delta}$
demonstrate that it displays Fermi liquid-like behavior \cite{mirzaei13},
indicating that the nature of the underdoped (pseudogap) region in the cuprate
materials is still controversial.

%
%
%
%

Interestingly, an almost identical behavior has also been observed in the $A$Fe$_2$As$_2$
($A=$ Ba and Ca) iron-arsenic compounds \cite{johnston10}.  In BaFe$_2$As$_2$ a
spin-density-wave (SDW) state develops below $T_{\rm SDW} \simeq 138$~K, resulting
in the formation of a Dirac-like cone in the electronic dispersion close to the
Fermi surface \cite{zabolotnyy09,richard10} with small pockets or puddles.  The
frequency-dependent scattering rate has a clear quadratic component just above
$T_{\rm SDW}$, suggesting the non-magnetic state of this material may be described as a
Fermi liquid (see Supplementary Fig.~S2a online); when the SDW transition is removed
by Co substitution, the quadratic behavior persists from 295~K down to 27~K (see
Supplementary Fig.~S2b online).
The optical conductivity of BaFe$_2$As$_2$ is shown in Fig.~1d; for $T \ll T_{\rm SDW}$
we once again observe the fractional power law in the residual low-frequency optical
conductivity \cite{hu08}, $\sigma_1 \propto \omega^{-1.2}$, similar to that seen in
the ground state of several of the cuprates.
The identical power law is also observed in CaFe$_2$As$_2$ for $T \ll T_{\rm SDW}$
(see Supplementary Fig.~S3 online).

%
%
\section*{Discussion}
One important aspect of the fractional power law lies in its ability to remove the nodal
quasiparticle (residual) response that masks the gap.
In a superconductor, the real part of the optical conductivity at low frequencies may
be expressed as the linear combination $\sigma_1(\omega) = \delta(0) + \sigma_{\rm qp} +
\sigma_{\rm gap}+\cdots$, where $\delta(0)$ is the zero-frequency component
that corresponds to the superfluid density, $\sigma_{\rm qp}$ is the conductivity
due to the unpaired quasiparticles, and $\sigma_{\rm gap}$ is the contribution due to direct
excitations across the gap.  [In the normal state, $\delta(0)$ is absent and $\sigma_{\rm qp}$
is just the quasiparticle response from the whole Fermi surface.]
Because we now have an explicit functional form for $\sigma_{\rm qp}$ for various
materials, then for $\omega>0$ we neglect $\delta(0)$ and the low-frequency response is
$\sigma_{\rm gap} \simeq \sigma_1(\omega) - \sigma_{\rm qp}$.
%
%
The conductivity due to the superconducting energy gap may be described
phenomenologically using a Kubo-Greenwood approach \cite{harrison} in which
all the zero-momentum transitions across the gap in the Brillouin
zone are considered; in general terms, the optical conductivity due to the gap
is a reflection of the joint density of states of the photo-excited electron and
hole pairs.  In a conventional superconductor with an isotropic energy gap $\Delta$
and weak coupling to phonons (or any other exchange boson), then for $T \ll T_c$ in
systems at or close to the dirty limit [$1/\tau_0 \gtrsim 2\Delta$ where $1/\tau_0 =
1/\tau(\omega \rightarrow 0)$], the onset of absorption will occur at $2\Delta$; for
modest coupling, this onset shifts to $\Omega_0+2\Delta$, where $\Omega_0$ is the
energy of the boson \cite{akis91,nicol91}.
Similarly, in a {\em d}-wave superconductor in the dirty limit with weak coupling,
the onset would be expected at $\omega\simeq 0$; however, for moderate coupling
the onset should shift to $\Omega_0$ with a local maximum at $\simeq \Omega_0+2\Delta_0$.

%
%
\begin{figure}
\centerline{\includegraphics[width=3.5in]{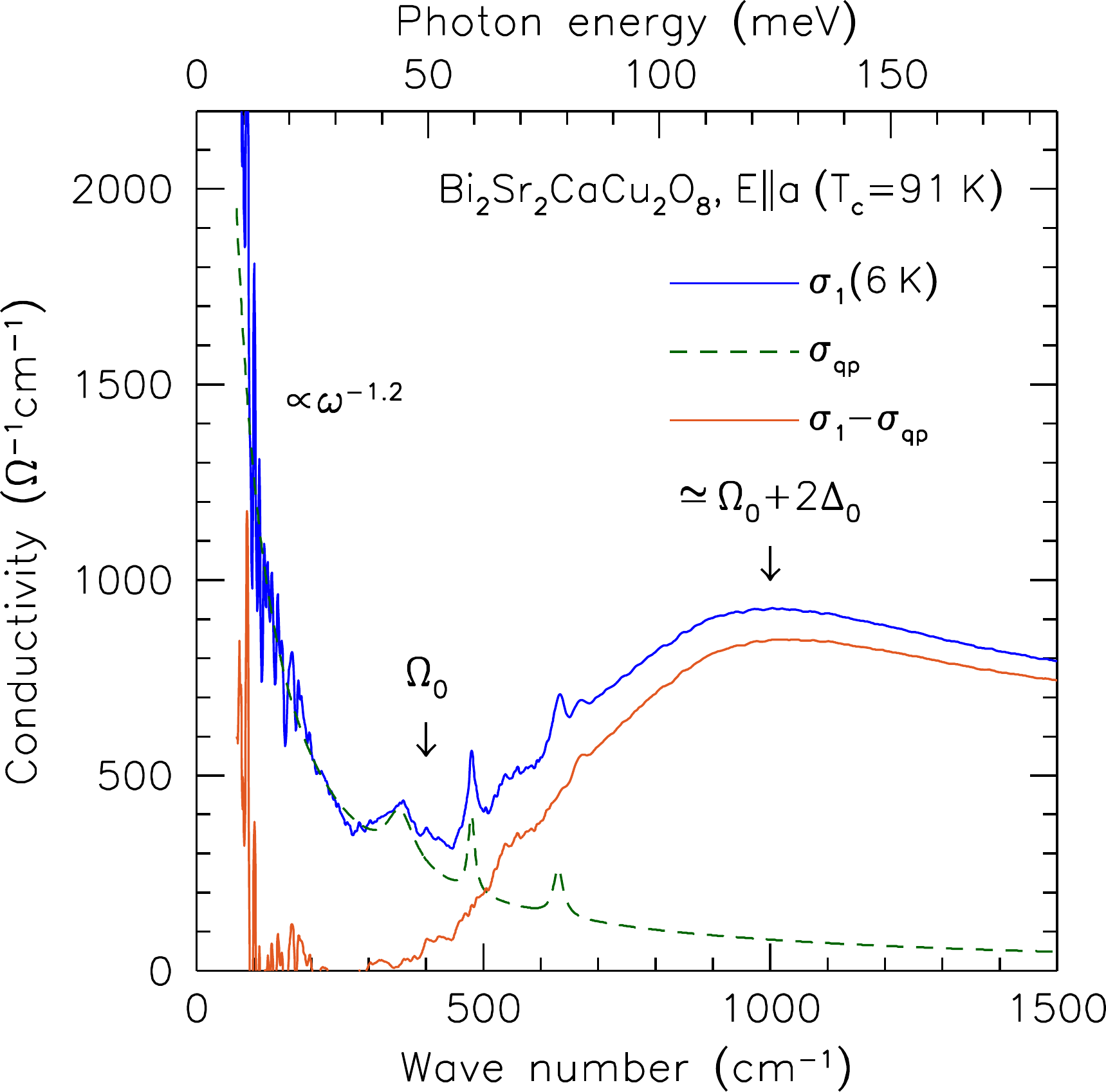}}%
\vspace*{0.4cm}
\begin{trivlist}
\item
\boldmath
\noindent{\bf\textsf{Figure 2} $|$ The decomposition of the optical conductivity in a cuprate superconductor.}
\unboldmath
The optical conductivity of optimally-doped Bi$_2$Sr$_2$CaCu$_2$O$_{8+\delta}$
at 6~K versus wave number (photon energy) with the residual quasiparticle conductivity
shown and removed; several sharp features in the conductivity have been fit to
Lorentzian oscillators (Supplementary Information) and have also been removed.
The subtracted spectra shows an onset of absorption at $\Omega_0$ and a local
maximum at $\simeq \Omega_0+2\Delta_0$.
\end{trivlist}
\end{figure}

%
%
The result for the removal of the quasiparticle contribution, $\sigma_1(\omega) -
\sigma_{\rm qp}$, is shown in Fig.~2 for Bi$_2$Sr$_2$CaCu$_2$O$_{8+\delta}$ at
$\sim 6$~K, well below $T_c$ the resulting conductivity is effectively zero at
low frequency and the onset of conductivity does not begin until $\omega \gtrsim
400$~cm$^{-1}$.  This corresponds to the bosonic excitation at $\Omega_0$,
the frequency above which a change occurs in the optical conductivity due to the
strong renormalization of the scattering rate (see Fig.~1a).
This indicates that there is at least moderate electron-boson coupling
in this material \cite{carbotte99,munzar99} and that the local maximum in the
conductivity will be at $\simeq \Omega_0+2\Delta_0$.
The inferred values of $\Omega_0 \simeq 50$~meV and $\Delta_0 \simeq 35$~meV are
in good agreement with estimates for these quantities based from angle-resolved
photoemission spectroscopy \cite{damascelli03}, and are consistent with optical inversion
techniques \cite{tu02,dordevic05}.  This procedure may also be successfully
applied to the YBa$_2$Cu$_3$O$_{6+y}$ materials (see Supplementary Fig.~S4
online), as well as the iron-based BaFe$_2$As$_2$ and CaFe$_2$As$_2$ materials in
their SDW states (see Supplementary Fig.~S5 online).

%
%
The significance of our finding is the common fractional power law behavior
of the low-frequency optical conductivity (THz and far-infrared regions) in
materials with Dirac cone-like electronic dispersion and nodal Fermi surfaces.
More generally, the fractional power law behavior signals the importance of
many-body effects in quantum materials with this unique electronic dispersion,
where the fractional power law in conductivity is roughly equivalent to a
nearly-linear frequency dependence of the scattering rate.
Similar results have been found in single layer graphene in the linear dependence
of the resistivity which is the result of electron-phonon (acoustic phonon)
coupling \cite{bolotin08}.  However, in the materials discussed here, the
electron-phonon coupling is weak.
The power-law behavior observed in this work is likely the result of the scattering
of nodal quasiparticles by low-energy (bosonic) excitations, or possibly some
unique self-energy effect of the Dirac-like quasiparticles.  What is common
for these systems are the existence of antiferromagnetic spin fluctuations
(or over-damped spin density waves in the SDW materials), which may be the
underlying mechanism that gives rise to the nearly linear frequency dependence
of the scattering rate.

\section*{Methods}
The temperature dependence of the absolute reflectance was measured at a
near-normal angle of incidence over a wide frequency range using an
{\em in situ} evaporation method \cite{homes93a}.  In this study
mirror-like as-grown crystal faces have been examined.
The complex optical properties were determined from a Kramers-Kronig analysis
of the reflectance \cite{dressel-book}.  The Kramers-Kronig transform requires
that the reflectance be determined for all frequencies, thus extrapolations
must be supplied in the $\omega \rightarrow 0, \infty$ limits.  In the metallic
state the low frequency extrapolation follows the Hagen-Rubens form,
$R(\omega) \propto 1-a\sqrt{\omega}$, while in the superconducting state
$R(\omega) \propto 1-a\,\omega^4$ is typically employed; however, it should
be noted that when the reflectance is close to unity the analysis is not
sensitive upon the choice of low-frequency extrapolation.  The reflectance
is assumed to be constant above the highest measured frequency point up
to $\simeq 1 \times 10^5$~cm$^{-1}$, above which a free electron gas
asymptotic reflectance extrapolation $R(\omega) \propto 1/\omega^4$ is
employed \cite{wooten}.

\section*{acknowledgments}
The authors would like to acknowledge useful discussions
with P. W. Anderson, Y. M. Dai, D. N. Basov, D. A. Bonn, S. V. Borisenko,
G. Kotliar, P. Phillips, J. D. Rameau, D. Schmeltzer and C. Varma.
Research supported by the U.S. Department of Energy, Office of
Basic Energy Sciences, Division of Materials Sciences and Engineering
under Contract No. DE-AC02-98CH10886.  C.C.H. would like to acknowledge
the hospitality of the Theory Institute for Strongly Correlated and
Complex Systems.

%
%

%
%

\section*{Supplementary Information}
\subsection*{Modeling the optical conductivity of the cuprates}

The interpretation of the free-carrier response in the low-frequency optical
conductivity of the high-temperature superconductors has always presented a
challenge due to the curious nature of the flat, essentially incoherent,
mid-infrared response \cite{basov05}.  Two approaches have been developed
to model this behavior.  The first is a single-component model which
considers delocalized carriers interacting strongly with an optically inactive (bosonic)
excitation in which the scattering rate and the effective mass both have a strong
frequency dependence \cite{allen77,puchkov96}.  The second approach is a two-component
model which associates the low-frequency conductivity with a free-carrier response,
and the mid-infrared conductivity with one or more bound excitations \cite{tanner}.

%
%
%
\begin{figure}[t]
%
%
\centerline{\includegraphics[width=3.2in]{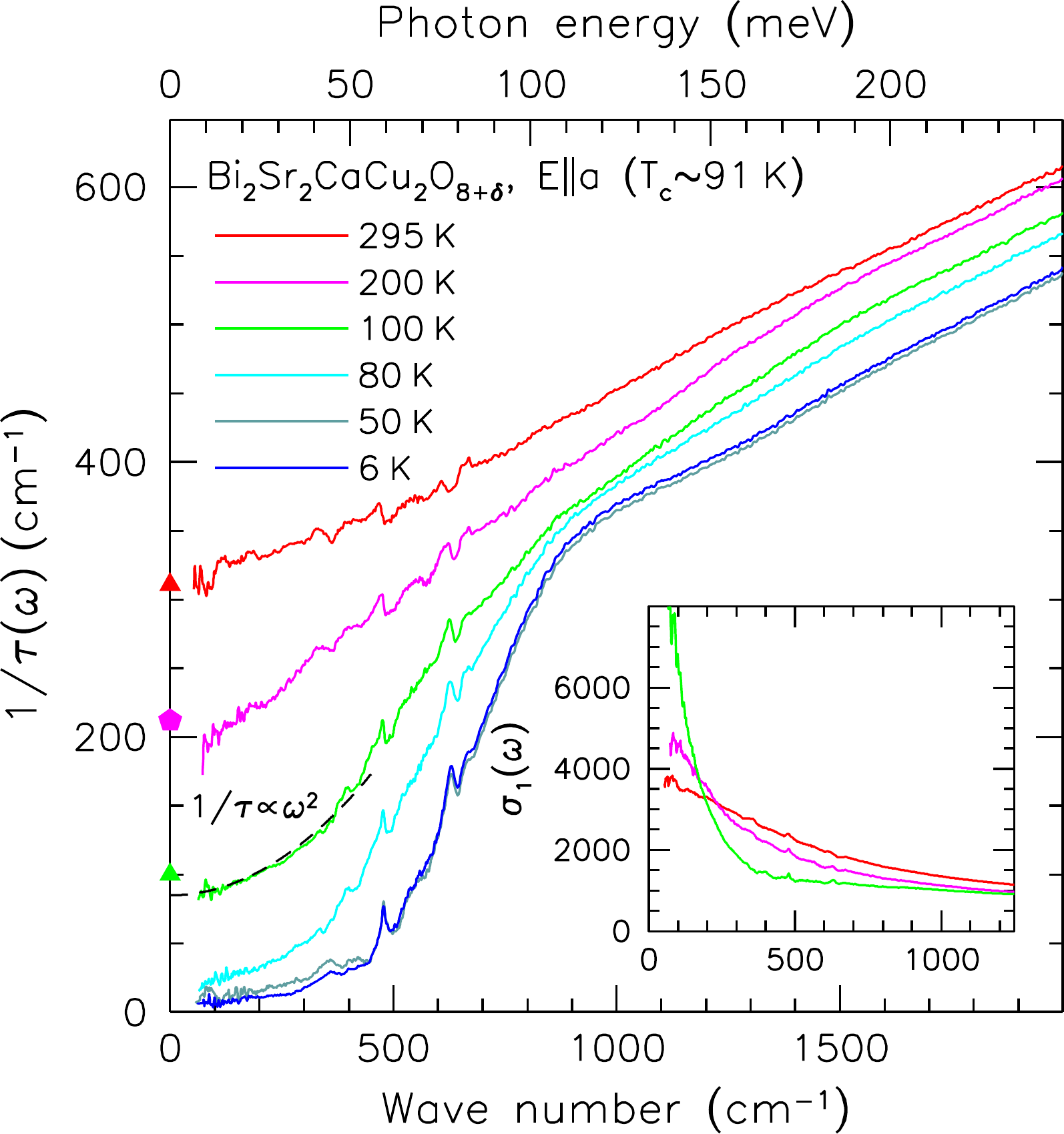}}%
\vspace*{0.4cm}
\begin{trivlist}
  \item
  \boldmath
{\bf\textsf{Figure S1} $|$ Frequency-dependent scattering rate of
Bi$_2$Sr$_2$CaCu$_2$O$_{8+\delta}$.}
\unboldmath
The temperature dependence of the frequency-dependent scattering rate versus
versus wave number (photon energy) for for the cuprate superconductor
Bi$_2$Sr$_2$CaCu$_2$O$_{8+\delta}$ at optimal doping ($T_c = 91$~K) for
light polarized in the {\em a-b} planes.  In the normal state, the scattering rates
determined from the Drude-Lorentz fits agree well the extrapolated $1/\tau(\omega
\rightarrow 0)$ values; however, below $T_c$ this model cannot accurately determine
the scattering rates of the residual quasiparticles.  Inset: the normal-state optical
conductivity.
\end{trivlist}
\end{figure}

%
%
Initially, we consider the two-component approach, where the optical
conductivity has been fit in both the normal state by assuming that the
scattering rate for the free carriers has a temperature dependence but is
frequency independent, $1/\tau(\omega,T) \simeq 1/\tau_0 + {\mathcal O}(T^2)$.
This is equivalent to the Drude model for a metal.  The mid-infrared conductivity
has been modeled with Lorentzian oscillators, so that the complex dielectric
function $\tilde\epsilon = \epsilon_1 + i\epsilon_2$ for these two contributions
is the simple Drude-Lorentz model
\begin{equation}
  \tilde\epsilon(\omega) = \epsilon_\infty - {{\omega_{p}^2}\over{\omega^2+i\omega/\tau}}
    + \sum_j {{\Omega_j^2}\over{\omega_j^2 - \omega^2 - i\omega\gamma_j}},
\end{equation}
where $\epsilon_\infty$ is the high-frequency contribution to the real part of
the dielectric function, $\omega_{p}^2=4\pi ne^2/m_b$ and $1/\tau$ are the square
of the plasma frequency and scattering rate, respectively, and $\omega_j$,
$\gamma_j$ and $\Omega_j$ are the position, width, and oscillator strength of
the $j$th vibration or bound excitation.  The complex conductivity is
$\tilde\sigma(\omega) = \sigma_1 + i\sigma_2 = i\omega\left[ \epsilon_\infty -
\tilde\epsilon(\omega) \right]/4\pi$, so that the real part of the optical conductivity
is $\sigma_1(\omega) = \omega\epsilon_2/60$ (in units of $\Omega^{-1}$cm$^{-1}$).
The Drude-Lorentz model has been fit to the real part of the optical
conductivity using a non-linear least-squares method.  The result for
the fit to optimally-doped Bi$_2$Sr$_2$CaCu$_2$O$_{8+\delta}$ ($T_c = 91$~K)
for light polarized along the {\em a} axis in the normal state at 100~K
yields $\omega_{p} \simeq 8560$~cm$^{-1}$ and $1/\tau \simeq 100$~cm$^{-1}$.
The agreement with experiment is quite good, but it should be noted that the
number and location of the Lorentzian oscillators required to achieve this fit
is somewhat arbitrary.  The presence of these mid-infrared oscillators is difficult
to physically justify given that the cuprates typically only have a single band
crossing the Fermi surface \cite{damascelli03}, with no other features nearby that
would readily account for these excitations.  This suggests that we should instead
consider a single channel, or single component, for the conductivity in which the
charge carriers are strongly renormalized.

%
%
%
The single-component model considers the simple Drude model for a uncorrelated metal,
and introduces a frequency dependence into the scattering rate and effective mass,
\begin{equation}
  \tilde\epsilon(\omega) = \epsilon_\infty - {{\omega_p^2}\over{
  \left[m^\ast(\omega)/m_b \right]\left[\omega^2+i\omega/\tau(\omega)\right]}}.
\end{equation}
The frequency-dependent scattering rate and mass enhancement factor may then
be determined experimentally \cite{allen77,puchkov96}:
\begin{equation}
  {{1}\over{\tau(\omega)}} = {{\omega_p^2}\over{4\pi}} {\rm Re}
  \left[ {{1}\over{\tilde\sigma(\omega)}} \right]
\end{equation}
and
\begin{equation}
  {{m^\ast(\omega)}\over{m_b}} =
  {{\omega_p^2}\over{4\pi}} {{1}\over{\omega}} {\rm Im}
  \left[ {{1}\over{\tilde\sigma(\omega)}} \right],
\end{equation}
where $m_b$ is the band mass, $m^\ast(\omega)/m_b = 1 + \lambda(\omega)$, and
$\lambda(\omega)$ is a frequency-dependent electron-boson coupling constant.
The temperature dependence of the frequency-dependent scattering rate for
optimally-doped Bi$_2$Sr$_2$\-CaCu$_2$O$_{8+\delta}$ is shown in Fig.~S1
using the previously determined value for $\omega_p$ in the normal state.
Aside from  $1/\tau(\omega)$ being somewhat smaller due to our
choice of $\omega_p$, it is consistent with other results \cite{hwang07}.
The polygons denote the values for $1/\tau$ returned from the Drude-Lorentz
fits.  In general, we would expect the fitted values for $1/\tau$ from the
Drude-Lorentz model to agree with $1/\tau(\omega\rightarrow 0)$ in the
normal state, and this is indeed the case.
At room temperature $1/\tau(\omega)$ is roughly linear in frequency, and therefore
consistent with a marginal Fermi liquid \cite{littlewood91,hwang04}.
As the temperature is lowered the scattering rate develops a quadratic
component just above $T_c$ up to $\simeq 400$~cm$^{-1}$, above which the scattering
rate increases dramatically, as reflected by the kink in the optical conductivity
at 100~K shown in the inset of Fig.~S1.  This suggests that just above $T_c$
the transport in the normal state may be cautiously described as a Fermi liquid;
this is in agreement with the proposed phase diagram of the high-temperature cuprate
superconducors \cite{broun08}.  Interestingly, a similar quadratic frequency
dependence in $1/\tau(\omega)$ has also been observed in the normal state of
the strongly underdoped single-layer cuprate material HgBa$_2$CuO$_{4+\delta}$
($T_c = 67$~K) and interpreted as evidence of Fermi-liquid like
behavior \cite{mirzaei13}, indicating that the underdoped region of the cuprates
is still a rapidly evolving area.
%
%
%
\begin{figure}[t]
\centerline{\includegraphics[width=3.2in]{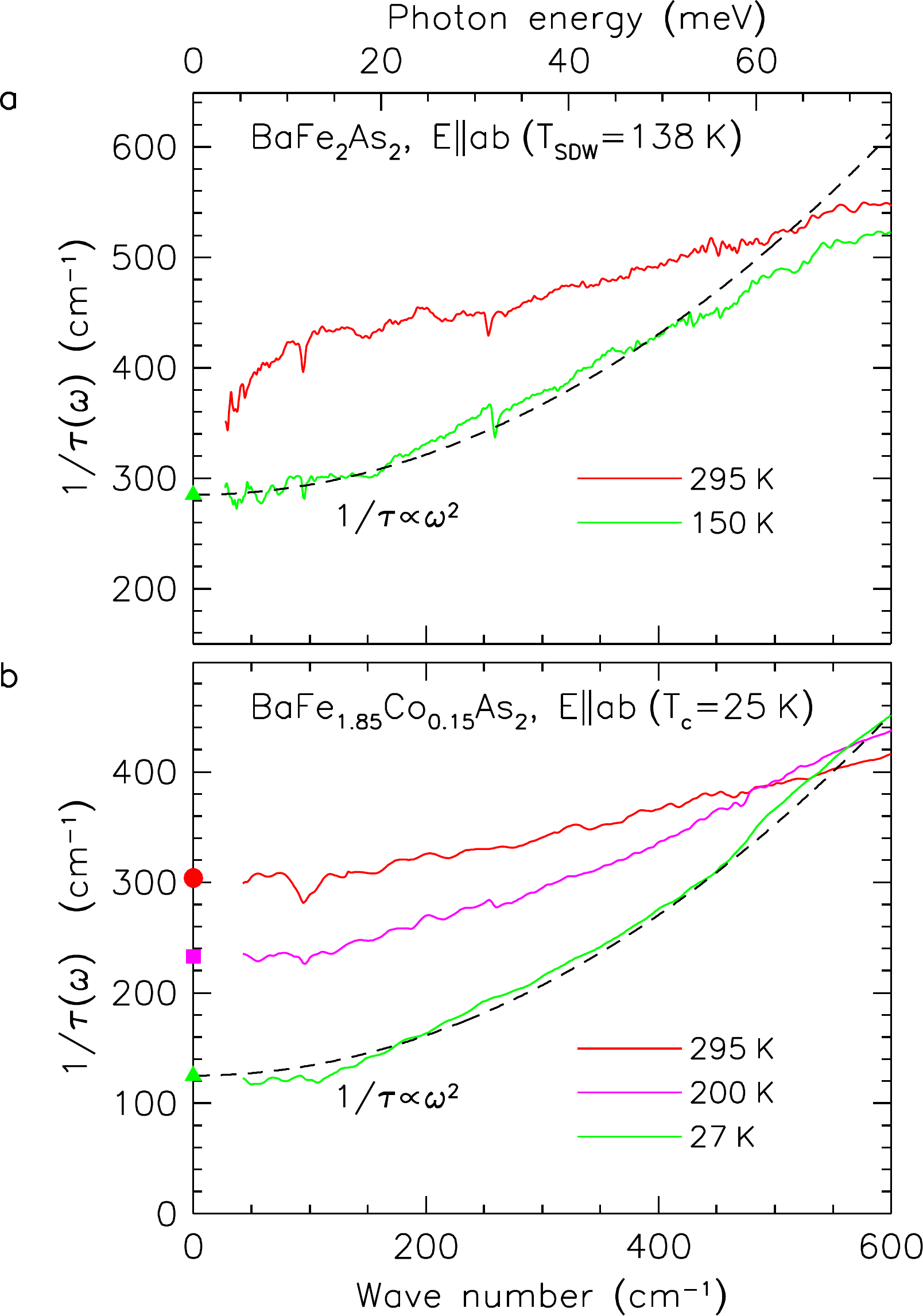}}%
\vspace*{0.4cm}
\begin{trivlist}
\item
\boldmath
\noindent{\bf\textsf{Figure S2} $|$ Frequency-dependent scattering rate
in iron-based mateials.}
\unboldmath
{\bf a},
The temperature dependence of the frequency-dependent scattering rate versus
versus wave number (photon energy) for BaFe$_2$As$_2$ for $T > T_{\rm SDW}$.
{\bf b}, The temperature dependence of the frequency-dependent scattering rate versus
versus wave number (photon energy) for BaFe$_{1.85}$Co$_{0.15}$As$_2$ for $T > T_{c}$.
In both cases a curve with a quadratic frequency dependence has been superimposed on
the low-temperature data.
\end{trivlist}
\end{figure}

Below $T_c$ the picture is considerably more complicated.  While the collapse of
the low-frequency scattering rate is consistent with the collapse of the
quasiparticle scattering rate observed in the microwave experiments \cite{shibauchi96},
it is in fact due to the formation of a condensate that dominates the
imaginary part of the optical conductivity, $\sigma_2$.  Adopting a simple
two-fluid model for the behavior of the superconducting state \cite{bonn93},
it may be shown that for $T \ll T_c$, the condition $\sigma_1 \ll \sigma_2$ is
satisfied; given $1/\tau(\omega) \propto \sigma_1/(\sigma_1^2+\sigma_2^2)$,
then to leading order $1/\tau(\omega) \propto \sigma_1/\sigma_2^2$,
indicating that below $T_c$ the scattering rate is dominated by the formation
of a superconducting condensate.  Therefore, in the superconducting state this
approach may not be used to make any reliable statements about the scattering
rate of the unpaired quasiparticles responsible for the residual conductivity.
%
%
The small values for $1/\tau$ for $T < T_c$ determined from microwave studies
indicate that we are in the $\omega\tau \gg 1$ regime, so that if the unpaired
quasiparticles constitute a Fermi liquid then $\sigma_1 \propto \omega^{-2}$
behavior should be recovered; however, we instead observe the fractional power
law $\sigma_1 \propto \omega^{-1.2}$.  This indicates that while the carriers
just above $T_c$ in the normal state might be described as a Fermi liquid,
the optical properties of the residual conductivity below $T_c$ describe a non-Fermi
liquid.  (We note that as $\omega\rightarrow 0$ we will eventually recover
$\omega\tau \sim 1$, at which point we would expect a roll-off in the conductivity
as it becomes largely frequency independent.)

%
%
In summary, the one and two-component models both provide useful information
about the normal and superconducting states.  However, in the Drude-Lorentz
picture the low-frequency Lorentzian oscillators required to obtain a reasonable
fit to the optical conductivity do not have a clear physical origin, and it
is likely that this is the result of the two-component model being forced to
deal with a strongly renormalized free-carrier scattering rate.  This suggests
that at low frequencies the single component model is a better description of the
delocalized carriers, which in the normal state close to $T_c$, may be cautiously
described as a Fermi liquid.

%
%
%
\begin{figure}[t]
\centerline{\includegraphics[width=3.2in]{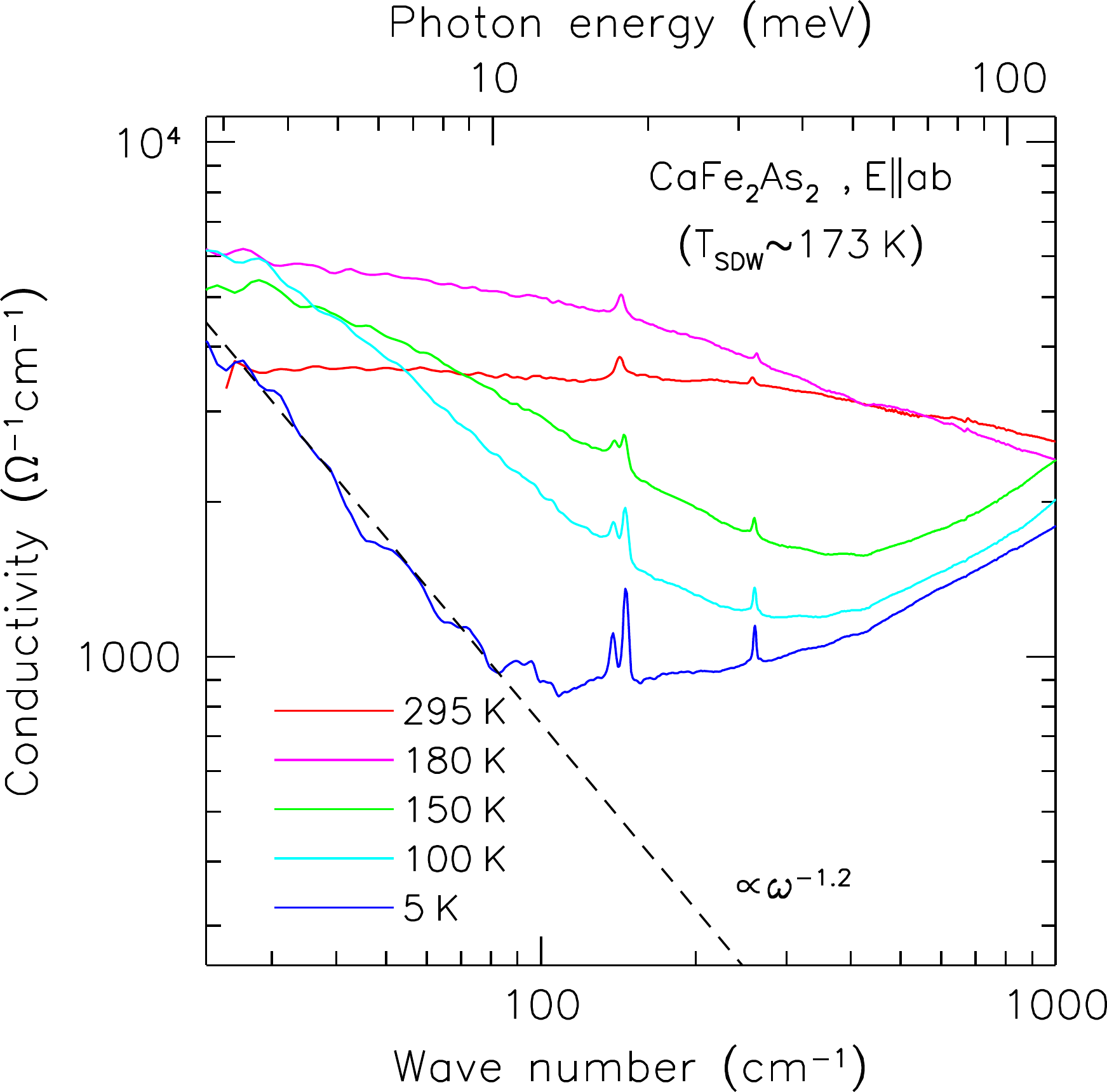}}%
\vspace*{0.4cm}
\begin{trivlist}
\item
\boldmath
\noindent{\bf\textsf{Figure S3} $|$ Fractional power law in the optical conductivity of CaFe$_2$As$_2$.}
\unboldmath
The log-log plot of the temperature dependence of the real part of the optical
conductivity versus wave number (photon energy) for the iron-based material
CaFe$_2$As$_2$ ($T_{\rm SDW} \simeq 173$~K) for light polarized in the {\em a-b}
planes.  The low-frequency data at 6~K displays a $\sigma_1 \propto \omega^{-1.2}$
dependence.
\end{trivlist}
\end{figure}

%
%
%
\subsection*{\boldmath BaFe$_2$As$_2$ and BaFe$_{1.85}$Co$_{0.15}$As$_2$ \unboldmath}
The pnictide materials are multiband systems \cite{johnston10} that have
been modeled using a two-Drude approach \cite{wu10} to describe the hole
and electron pockets.  However, in some of these materials it may be
argued that a single band dominates the transport properties; the
assumption of a single band allows the cautious application of the
generalized Drude model.
The frequency dependent scattering rate of the pnictide BaFe$_2$As$_2$
($T_{SDW} = 138$~K) develops a clear quadratic frequency dependence in
the non-magnetic state for $T \gtrsim T_{SDW}$, shown in Fig.~S2a,
suggesting that this material may be described as a Fermi liquid.
Interestingly, when the magnetic and structural transition in this material is
suppressed though Co substitution in BaFe$_{1.85}$Co$_{0.15}$As$_2$ and
superconductivity is induced ($T_c = 25$~K), the quadratic behavior is observed
in the normal state at low temperature, $T \gtrsim T_c$, shown in Fig.~S2b.

%
%
%
\subsection*{\boldmath CaFe$_2$As$_2$ \unboldmath}
The iron-arsenic CaFe$_2$As$_2$ material is similar to BaFe$_2$As$_2$, having
a spin-density-wave (SDW) transition at $T_{\rm SDW} \simeq 173$~K, below which
it remains metallic.  The non-Fermi liquid power law behavior observed in the
low-frequency optical conductivity in the ground state of the cuprate materials
and BaFe$_2$As$_2$, $\sigma_1 \propto \omega^{-1.2}$, is also observed in
CaFe$_2$As$_2$ material well below $T_{\rm SDW}$, as shown in Fig.~S3.

%
%
%
\begin{figure}[t]
\centerline{\includegraphics[width=3.35in]{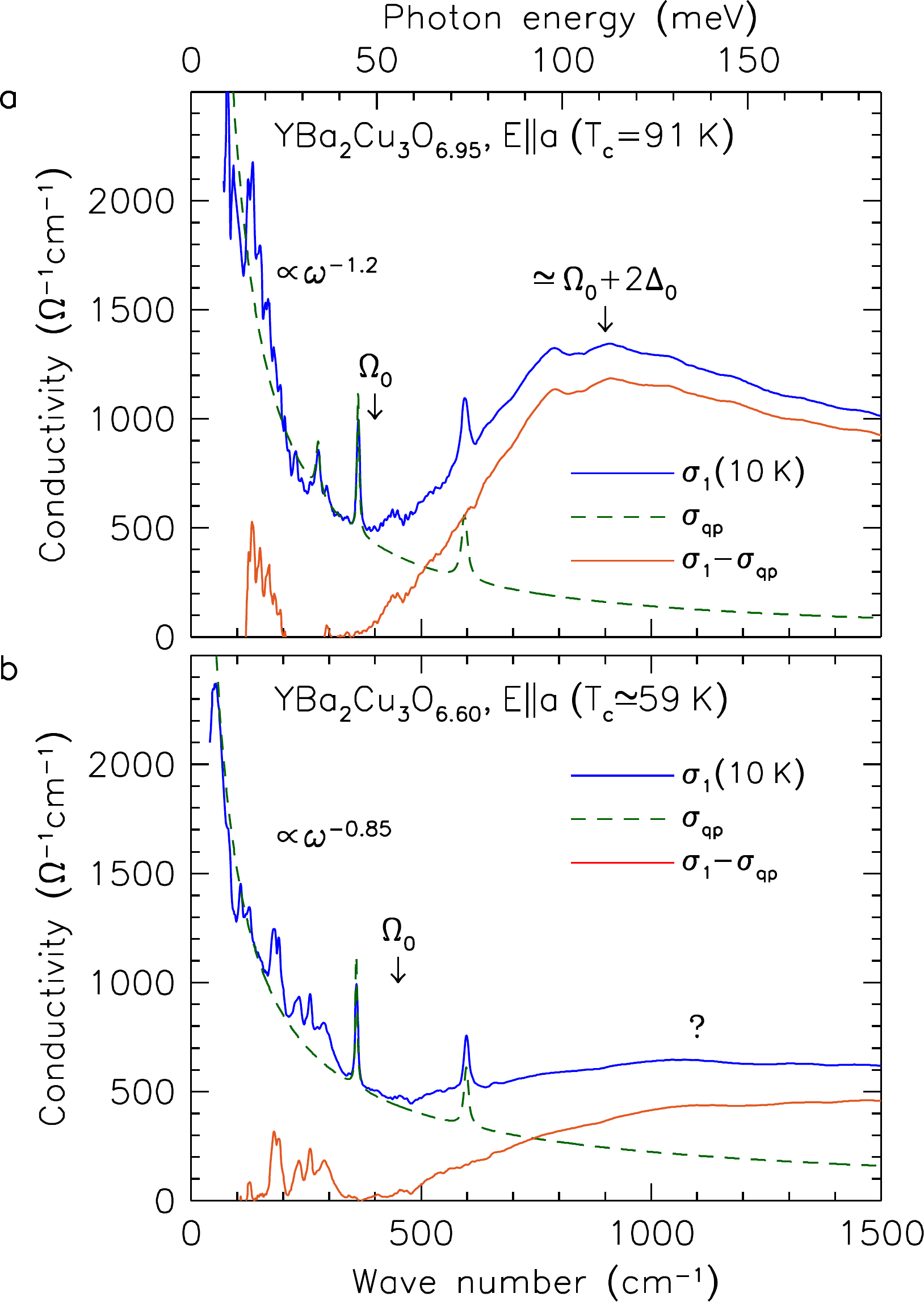}}%
\vspace*{0.4cm}
\begin{trivlist}
\item
\boldmath
\noindent{\bf\textsf{Figure S4} $|$ The decomposition of the optical conductivity of
YBa$_2$Cu$_3$O$_{6+y}$.}
\unboldmath
{\bf a}, The optical conductivity of optimally-doped YBa$_2$Cu$_3$O$_{6.95}$
versus wave number (photon energy) at 10~K with the residual quasiparticle
conductivity removed; the subtracted spectra shows an onset of absorption at
$\Omega_0$ and a local maximum at $\Omega_0+2\Delta_0$.
{\bf b}, Underdoped YBa$_2$Cu$_3$O$_{6.60}$ at 10~K.
\end{trivlist}
\end{figure}

%
%
\subsection*{\boldmath Decomposition of the optical conductivity of YBa$_2$Cu$_3$O$_{6+y}$ \unboldmath}
The sharp structures in the conductivity in optimally-doped YBa$_2$Cu$_3$O$_{6.95}$,
shown in Fig.~S4a, and in underdoped YBa$_2$Cu$_3$O$_{6.60}$, shown in Fig. S4b,
for light polarized along the crystallographic {\em a} axis are attributed to the
normally infrared-active $B_{3u}$ modes \cite{homes00}, which have been fit to
the data using Lorentzian oscillators (described earlier),
in linear combination with a polynomial background.  The oscillators
are added to the fractional power law expression for the conductivity for the
unpaired quasiparticles below $T_c$ and subtracted from the optical conductivity.
In both cases a value of $\Omega_0 \simeq 50$~meV is obtained, which is once
again consistent with inversion methods \cite{dordevic05}.  In the
optimally-doped material, the estimate of $\Delta_0 \simeq 30$~meV is similar
to what is obtained in tunneling studies \cite{yeh01}. However, in the
underdoped data there is a considerable amount of structure at low
frequency and there is no longer an obvious feature that can be identified
with the gap maximum.

%
%
\begin{figure}[t]
\centerline{\includegraphics[width=3.2in]{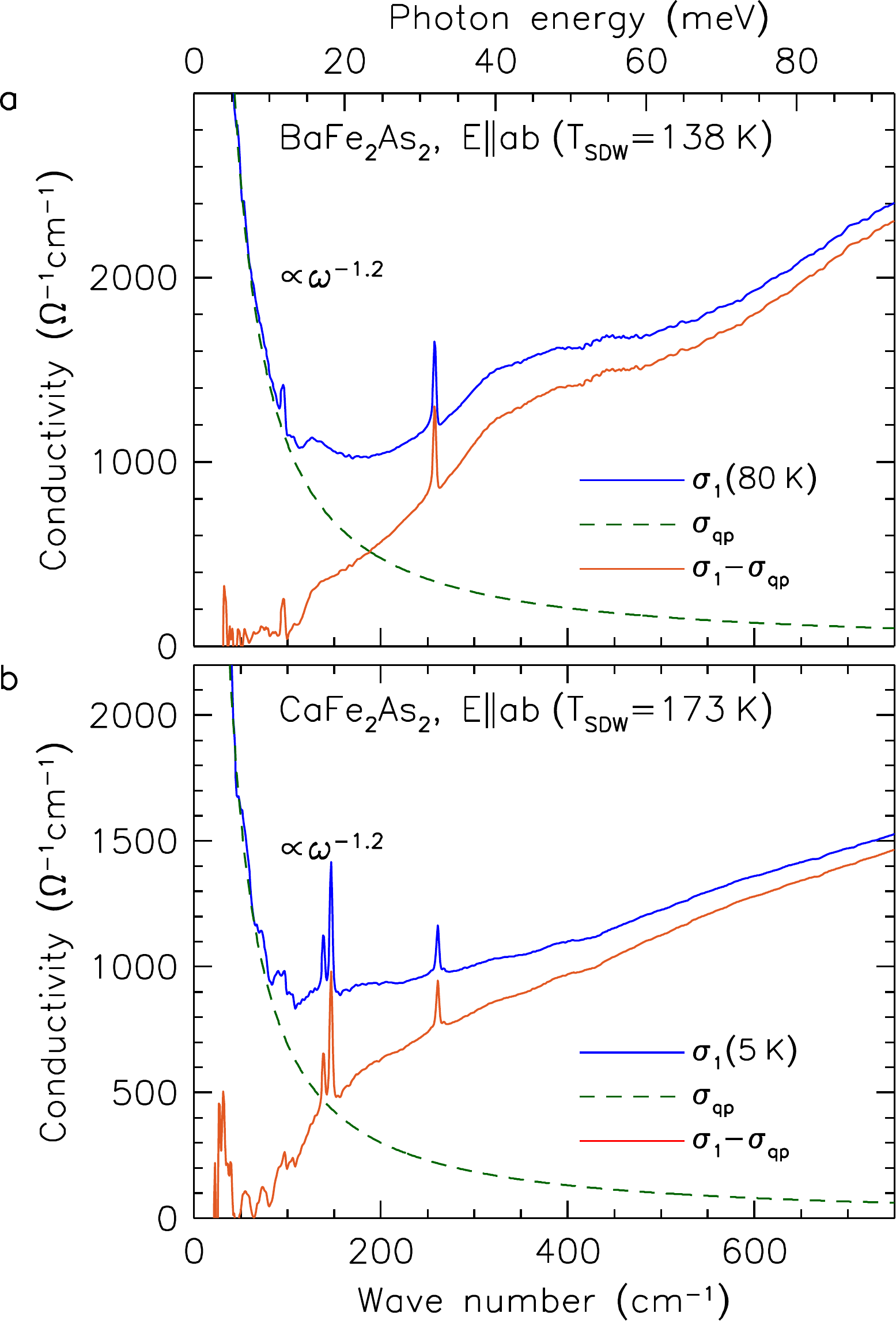}}%
\vspace*{0.4cm}
\begin{trivlist}
\item
\boldmath
\noindent{\bf\textsf{Figure S5} $|$ The decomposition of the optical conductivity
in iron-based mateials.}
\unboldmath
{\bf a},
The optical conductivity of BaFe$_2$As$_2$ versus wave number (photon energy) at
80~K with the residual quasiparticle conductivity removed showing the onset of
absorption at $\omega \gtrsim 0$.
{\bf b}, CaFe$_2$As$_2$ at 5~K.
\end{trivlist}
\end{figure}

%
%
\subsection*{\boldmath Decomposition of the optical conductivity of 
  BaFe$_2$As$_2$ and CaFe$_2$As$_2$ \unboldmath}
In the iron-pnictide materials, the metallic compounds BaFe$_2$As$_2$ and
CaFe$_2$As$_2$ have magnetic and structural transitions at $T_{\rm SDW} = 138$
and 173~K, respectively, below which they remain metallic \cite{johnston10}.
Below $T_{\rm SDW}$ the Fermi surface undergoes a reconstruction and a Dirac-cone
like gap opens in the electronic dispersion at (or close to) the Fermi
level \cite{richard10,sugai12}.  In BaFe$_2$As$_2$ at 80~K, and CaFe$_2$As$_2$ at
5~K, the low-frequency conductivity is described quite well by the fractional
power law $\sigma_1 \propto \omega^{-1.2}$.  The residual conductivity has been
subtracted from the optical conductivity in these materials in Fig.~S5 revealing
that, unlike the cuprate superconductors surveyed where the onset of absorption
does not begin until $\sim \Omega_0$, the absorption commences at a much lower
frequency, $\omega \gtrsim 0$.

%
%
%
%
%
%

\end{document}